\documentclass[9pt,twocolumn,twoside]{HCNB-preprint}
\setarticletype{Pre-print}
\usepackage[utf8]{inputenc}
\usepackage{lineno}
\usepackage{orcidlink}
\usepackage{xcolor}
\usepackage{siunitx}

\title{Minimal-footprint photonic crystal nanolasers for biointegration}
\author[1]{Catriona A. Thomson \orcidlink{0000-0003-2226-0586}}
\author[1]{Andreas Stühler}
\author[1]{Nachiket Pathak \orcidlink{0000-0001-6735-2332}}
\author[1]{Valeryia Dzikevich}
\author[1,*]{Marcel Schubert \orcidlink{0000-0002-8739-4852}}
\affil[1]{Humboldt Centre for Nano- and Biophotonics, Institut for Light and Matter, Department of Chemistry and Biochemistry, University of Cologne, Greinstr. 4-6, 50939 Cologne, Germany}
\affil[*]{marcel.schubert@uni-koeln.de}

\begin{document}

\begin{abstract}
Photonic crystals allow unprecedented control over how light is confined, propagates, and interacts with matter. Their development has had a transformative impact on optics and physics, and they remain the central platform for both fundamental discoveries and practical photonic technologies. However, the relatively large footprint and substrate-bound nature of photonic crystal structures have so far strongly limited their use as miniature optical devices or biointegrated sensors. Here, we overcome these limitations by identifying the minimal size of a 2D photonic crystal array needed to achieve lasing and describe the fabrication of substrate-less hexagonal laser particles with an active area as small as \SI{30}{\micro\meter\squared}. Massively parallel fabrication, robust detachment, and integration of the nanolaser particles into live cells is demonstrated. Crucially, by engineering spatial and spectral mode characteristics, we designed NIR-II probes with mode volumes on the order of tens of attolitres, an order of magnitude smaller than whispering gallery probes of similar dimensions. Such high light localization is comparable in scale to different organelles of eucaryotic cells. In the future, we expect that chemical or plasmonic functionalization of the device will enable label-free sensing of nanoscale intracellular processes, and that it shall serve as a miniature platform to exploit developments in optical and quantum sensing for chemical and biological applications.
\end{abstract}

\maketitle

\section{Introduction}
\begin{figure*}[h!]
\centering
\includegraphics[width=0.9\linewidth]{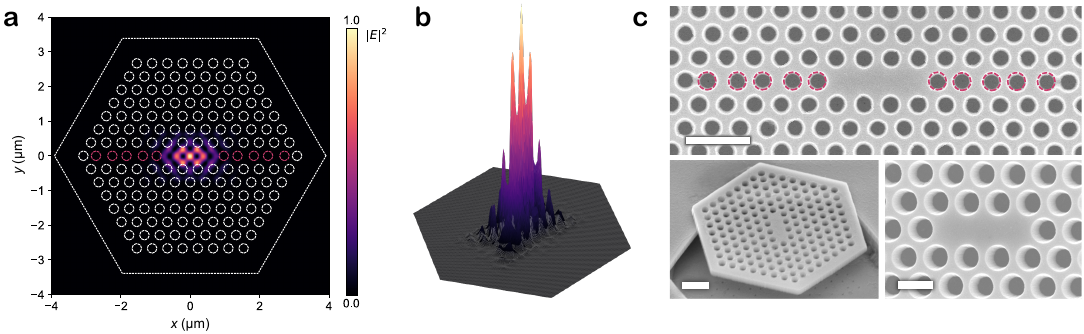}
\caption{\textbf{A small-footprint L3 photonic crystal nanolaser particle of InGaAsP:} \textbf{(a)} $|E|^2$ distribution of the fundamental L3 mode of the optimized design on a finite-sized hexagonal particle. The five perturbed holes are highlighted in red. \textbf{(b)} 3D representation of the mode shown in (a), highlighting the strong localization of the field inside the defect area. \textbf{(c)} SEM images of fabricated nanolaser particles. \textit{Top:} top-view of an optimized L3 cavity, showing the accurate placement of the perturbed holes (scale bar: \SI{100}{\nano\meter}). \textit{Bottom-left:} Detached optimized L3 nanolaser particle resting on the substrate after ultra-sonication (scale bar: \SI{500}{\nano\meter}). \textit{Bottom-right:} Titled view of the defect area (scale bar: \SI{500}{\nano\meter}).}
\label{fig:overview}
\end{figure*}

\begin{figure*}[h!]
\centering
\includegraphics[width=0.85\linewidth]{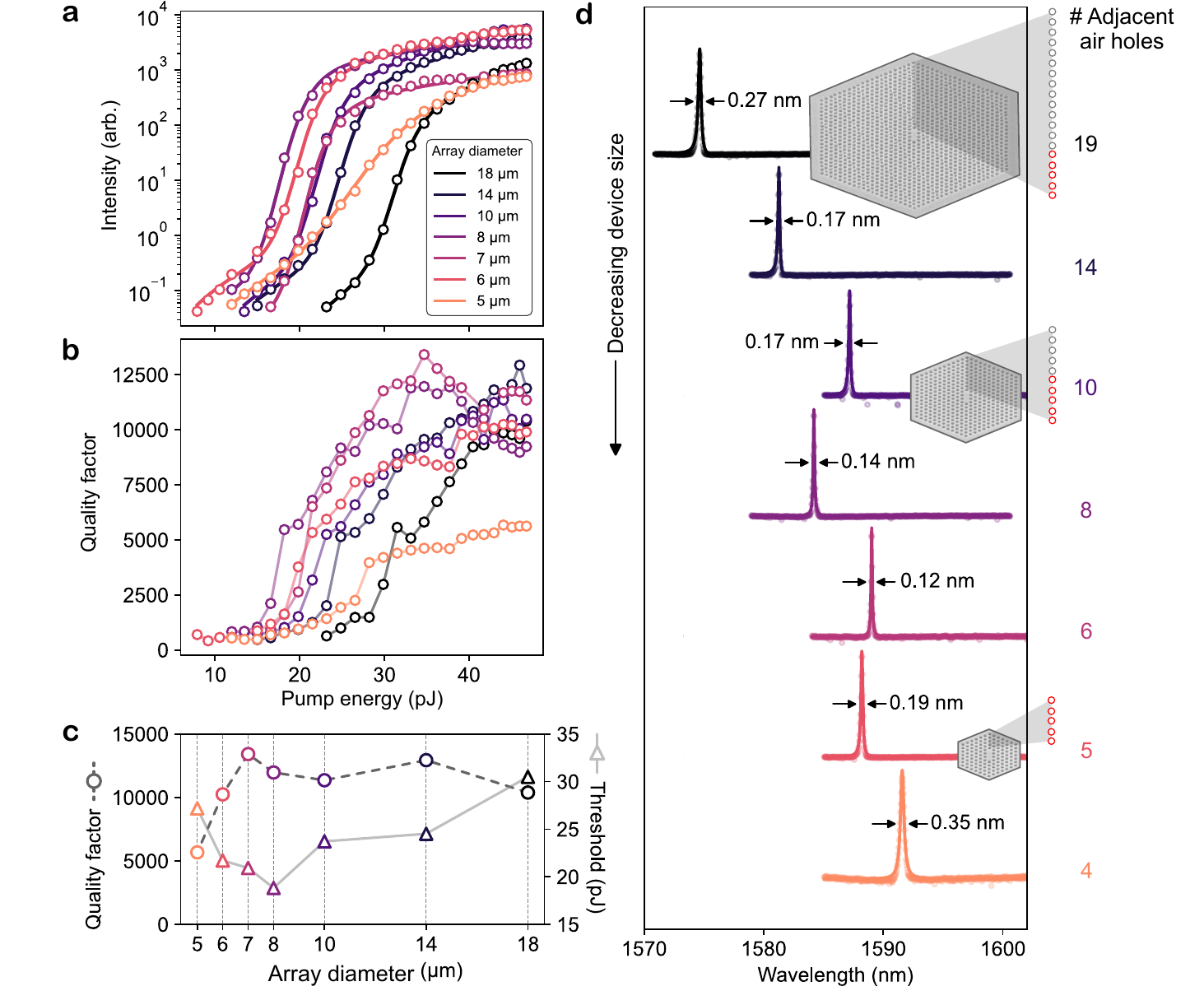}
\caption{\textbf{Determining the minimum array size for the optimized L3 cavity array nanolaser particles.} \textbf{(a)} Threshold curves of nanolaser particles showing the emitted power as function of the pump energy for PhC cavities of various sizes. \textbf{(b)} Calculated quality factors corresponding to the threshold curves in (a). \textbf{(c)} Summary of laser threshold energies (triangles) and maximum quality factors (circles) plotted over the array diameter. \textbf{(d)} Lasing spectra at a pump energy of \SI{34.7}{\pico\joule} for nanolaser particles with increasing array diameter. Measured SEM images of three nanolasers are overlaid for size comparison, where the array of adjacent holes (optimized positions highlighted in red) is schematically shown. Colours correspond to the previous panels. For each nanolaser size, the number of air holes adjacent to one side of the L3 defect is also stated. Arrays with less than 4 adjacent holes did not show lasing.}
\label{fig:devicesize}
\end{figure*}

\begin{figure*}[h!]
\centering
\includegraphics[width=0.9\linewidth]{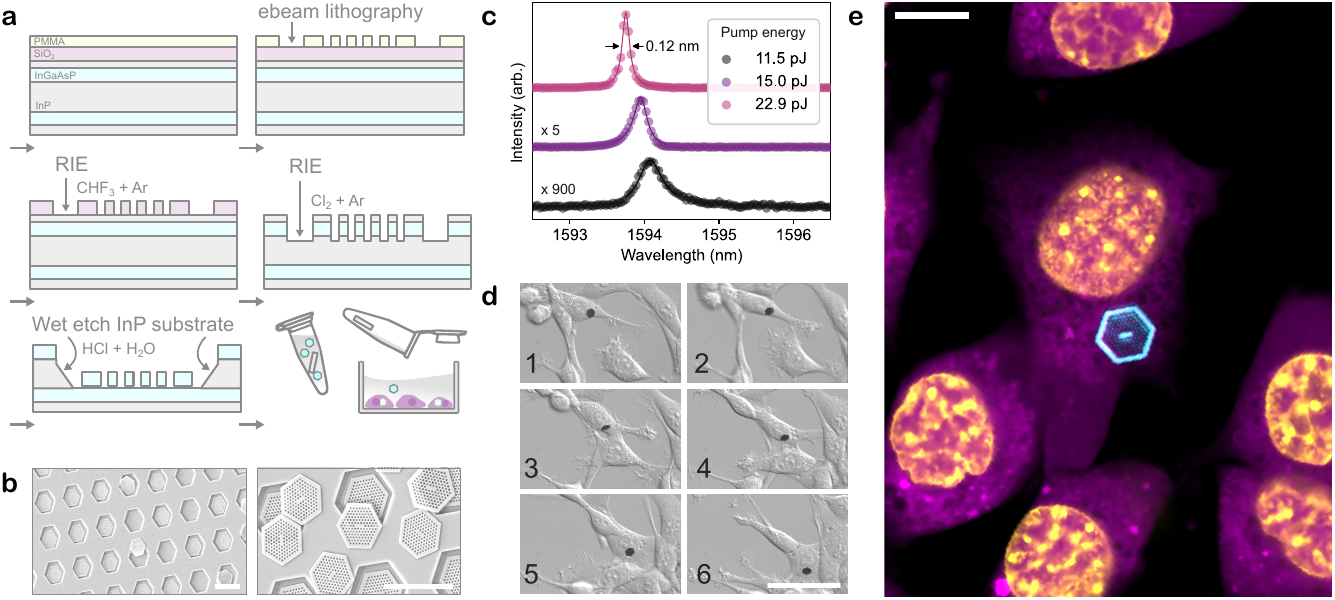}
\caption{\textbf{L3 nanolaser particles inside live fibroblast cells.} \textbf{(a)} Schematic overview of the procedures used to fabricate detached nanolaser particles and their addition to live cell cultures. First, the hole array and trench around each device are defined by e-beam lithography. After dry-etching through all wafer layers, wet-etching the InP substrate layer with hydrochloric acid allows each device to fall onto the etch-stop layer (also InGaAsP). Ultrasonication in PBS detaches the nanolaser particles into solution. Replacing PBS with cell media, the solution is added to cells and incubated for 24 hours to allow for nanolaser internalization. \textbf{(b)} SEM images of nanolasers after wet-etching and brief ultrasonication step (left). Scale bar, \SI{10}{\micro\meter}. Longer ultrasonication detaches almost all nanolaser particles (right). Scale bar, \SI{10}{\micro\meter}. \textbf{(c)} Normalized lasing spectra from an optimized L3 nanolaser inside a live fibroblast cell, measured at three different pump energies. \textbf{(d)} Brightfield microscopy time-lapse imaging of fibroblast cells and an intracellular nanolaser, covering a period of 4 hours. Scale bar, \SI{100}{\micro\meter}. \textbf{(e)} Confocal fluorescence microscopy image of fixed fibroblasts with stained cytosol (purple), nucleus (orange), and a L3 nanolaser (cyan). Scale bar, \SI{10}{\micro\meter}.}
\label{fig:internalized}
\end{figure*}
Photonic crystals (PhCs) are among the most advanced and versatile optical technologies currently available. They have been instrumental in understanding the similarities between light and matter waves by representing a photonic version of the periodic structure of ordered atoms inside crystalline materials. Many other directions of modern physics have used PhCs as a platform to study the related optical analogue, including phenomena related to wave chaos or topological science.

Two-dimensional semiconductor slabs patterned with circular, triangular or square air-holes are the basis for the majority of functional PhC devices performing in the visible to terahertz domain. Creative design choices based on the superposition and perturbation of PhC lattices -- leveraging intuition, topological theory, and inverse design algorithms~\cite{Jiang2020} -- enable high control over the resultant photonic modes confined to the dielectric slab. In recent years there have been significant advances in the development of devices exhibiting optimized out-coupling~\cite{Minkov2017,Portalupi2010}, fabrication-error tolerance~\cite{Haagen2025}, multi-mode operation~\cite{Tian2025}, subdiffraction limited optical field volumes, and exceptionally high quality factors (Q-factors)~\cite{Ouyang2024,Ma2023,Mao2021,Minkov2014,Minkov2017,Wang2025a}.

PhC light confinement and guiding has been exploited for cavity quantum electrodynamics~\cite{Yan2025}, on-chip optical communications~\cite{Notomi2004,Nozaki2010}, and low-threshold lasing, with one of the most practical applications being in extremely powerful electrically-driven lasers for commercial and clinical applications~\cite{Noda2024,Kurosaka2010,Yin2025}. Harnessing the strong light-matter coupling offered by their small mode-volumes and high Q-factors, chemical and biosensing with PhCs is also an active field, based on resonant wavelength shifts due to refractive index changes in the evanescent field~\cite{Inan2017,Lee2007,DiFalco2009,Scullion2011,Pitruzzello2018,Watanabe2022,Hamed2025,Khedr2025}, with some cavities including surface biofunctionalization or plasmonic elements for further mode confinement~\cite{Liang2017,DeAngelis2008,Xavier2017}.

Given their high configurability and potential for mass production, PhC cavities are well suited to the ambitious lab-on-chip vision, to interrogate multiple or complex biological signals on an ultra-compact system for the detection of disease~\cite{Watanabe2022,Harikrishnan2025,Subramanian2018}. However, as most PhC devices are confined to the substrate plane, there has been limited demonstration of direct interfacing with the cellular environment, with the most promising demonstration being a photonic crystal needle probe which achieved \textit{in vitro} label-free protein sensing inside a single cell~\cite{Shambat2013}.

Alternative micro and nanoscopic optical devices have been used for complete biointegration down to the single cell level, arguably enabled by their significantly simplified structures and fabrication. For example, intracellular microlasers represent a fast developing technology that aims to generate bright and spectrally narrow signals to encode real-time chemical and physical information about the cellular environment, and to serve as cellular barcodes~\cite{Thomson2025,Martino2019,Dannenberg2021,Kwok2023,Fikouras2018,Titze2022,Titze2024}. First demonstrations of intracellular lasers used spherical resonators to support whispering gallery modes (WGMs) oscillating just within the resonator’s perimeter~\cite{Schubert2015,Humar2015}. In addition to microspheres, microdisks made from III-V semiconductor gain materials (e.g. GaInP, InGaAsP, InP~\cite{Fikouras2018,Martino2019,Cho2025}) have become a popular platform for WGM microlasers.


WGM microlasers have also demonstrated various forms of biosensing including sensing protein binding~\cite{Caixeiro2023}, DNA hybridization~\cite{Caixeiro2025} and cardiac contractility~\cite{Schubert2020}. While WGM microlasers benefit from facile fabrication, their extended WGM severely limits spatially localized sensing and introduces perturbances of the lasing modes especially inside the highly dynamic and inhomogeneous environment of living cells. 

In this work, we demonstrate the miniaturization of two-dimensional photonic crystal nanolasers with the aim to be fully detached from the supporting substrate and to be internalized by living cells. The presented device is a high quality-factor, minimal-footprint InGaAsP L3 (3-point linear defect) PhC nanolaser particle with a mode volume of 0.76 ($\lambda$ / $n$)$^3$ and a lasing wavelength of 1580 nm. Crucially, the PhC mode supports a sensing volume on the order of tens of attolitres, an order of magnitude smaller than whispering gallery probes of similar dimensions. Such high light localization, potentially combined with chemical or plasmonic device functionalization, is expected to enable label-free sensing of nanoscale intracellular phenomena perhaps down to the single-molecule level.

\begin{figure*}[ht!]
\centering
\includegraphics[width=\linewidth]{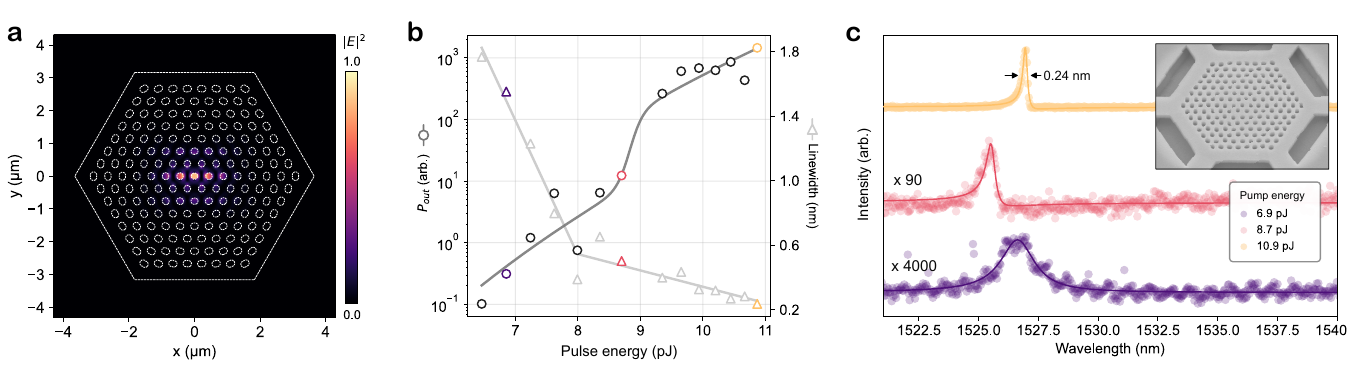}
\caption{\textbf{Nanolaser particles with alternative cavity design.} \textbf{(a)} Simulated $|E|^2$ distribution of the dipole mode of a hexagonal particle with a twisted lattice cavity (\SI{1.25}{\degree} twist between regular hexagonal lattices). \textbf{(b)} Lasing characteristics of a nanolaser fabricated from InGaAsP, showing the emitted power (triangles, left axis) and linewidth of the laser mode (circles, right axis) as function of the pump energy. \textbf{(c)} Lasing spectra taken from (b) for the twisted cavity nanolaser below (purple), at (red), and above (yellow) the laser threshold. \textbf{(Inset)} Tilted SEM image of the measured twisted cavity. Scale bar, 500 nm.}
\label{fig:twisted}
\end{figure*}
\section{Photonic crystal nanolasers with minimal footprint}
PhCs often require relatively large-scale (> $50 \times 50$~\unit{\micro\meter\squared}) lattices to control and manipulate light to enable e.g. wave-guiding, high-quality factor lasing, or sensing. Comparatively little attention has been given to miniature photonic crystal arrays until very recently~\cite{Ma2023,Wang2018,Wang2025a,Ouyang2024}. However, successful biointegrability and targeted microscopic sensing demand that the nanolasers are made with the smallest dimensions possible. We therefore assess the limits of miniaturization by producing an ensemble of cavities with decreasing footprints.

The cavity design used to start the miniaturization process is the quintessential L3 cavity consisting of a triangular lattice of air holes with three holes missing from the center (see Fig.~\ref{fig:overview}). For this cavity, minor displacements of holes adjacent to the defect zone are known to produce dramatically improved quality factors, which may be understood as the manipulation of the electric field distribution to avoid lossy interactions with hole interfaces~\cite{Yin2025}. We also reason that by decreasing the cavity size the decrease in quality factor would ultimately set the limit on the minimum dimensions of the nanolaser. To this end, the positions of the five holes directly adjacent to the defect were optimized for maximal quality factor. For simplicity, we started with a hole configuration found in an L3 cavity optimization by Minkov \textit{et~al.}~\cite{Minkov2014} for a silicon slab which achieved a simulated quality factor of 5 million. To adapt the design to our particular slab thickness, material and desired resonant wavelength, we performed simulations using a grid-search method to tweak the optimal hole positions in three iterations (see Ext. Data Fig.~\ref{fig:optimization}).

To determine the impact of truncating the PhC lattice into a small particle nanolaser, we produced on-chip cavities, with between 3 and 19 holes adjacent to the defect zone in both directions (see Fig.~\ref{fig:devicesize}). The corresponding \textit{array diameter} describes $2 \times$ the side-length of a regular hexagon within which all hole centers reside. The nanolasers were fabricated from a 240~nm thick slab of InGaAsP ($n = 3.445$). A 100~nm hard mask of SiO$_2$ was first sputtered onto the InP cladding, before spin-coating 230~nm of PMMA electron beam resist. Each hexagonal device was designed with supportive struts surrounding it to create an air-bridge cavity. Optimized photonic crystal arrays with a lattice spacing of 440~nm and a hole size of $0.32a$ were defined in the resist via e-beam lithography, before inductively-coupled plasma reactive-ion etching (ICP RIE) of the hard mask and semiconductor layers with CHF$_3$ and Cl$_2$ chemistries respectively. To produce suspended PhC cavities, the InP substrate was etched with diluted hydrochloric acid (see Fig.~\ref{fig:internalized}a). The chip was then submerged in phosphate buffered saline (PBS) solution during the optical characterization which resembles the anticipated aqueous target environment of the nanolasers.

To optically characterize the devices of various sizes, a \SI{1064}{\nano\meter} pulsed laser pump (5 ns, 2 MHz) was focused on the centre of the device of interest, located on the sample plane of an inverted microscope with a 100x NIR objective. The emission from the PhC cavities was collected through the same objective and analyzed on a spectrometer coupled to a cooled NIR InGaAs camera, with a spectral resolution of approximately 120~pm.

Fig.~\ref{fig:devicesize} shows the lasing performance for devices produced with four or more adjacent hole pairs, equivalent to array diameters between \SI{5}{\micro\meter} and \SI{18}{\micro\meter}. Smaller arrays were found to only exhibit spontaneous emission filtered by the cavity resonance. For all lasing devices, clear thresholds were observed, between 15-35~pJ, with corresponding linewidth-narrowing. Interestingly, the device which showed the highest quality factor of over 12,500 was the \SI{7}{\micro\meter} cavity. We also observe that the nanolasers with intermediate array sizes show the highest emission intensities and lowest thresholds. Lasing finally disappeared completely at an array size of \SI{4}{\micro\meter}. In addition, we analysed a previously described method where only one hole is shifted on an array with intermediate size and also found an optimized array with lower thresholds, higher quality factor, and increased brightness (see Ext. Data Fig.~\ref{fig:twoshifted}). Overall, these results demonstrate that it is possible to dramatically decrease the PhC size while maintaining high quality factors and lasing. In a next step we therefore detached the nanolasers from the wafer and investigated their performance inside living cells.

\begin{figure*}[h!]
\centering
\includegraphics[width=0.9\linewidth]{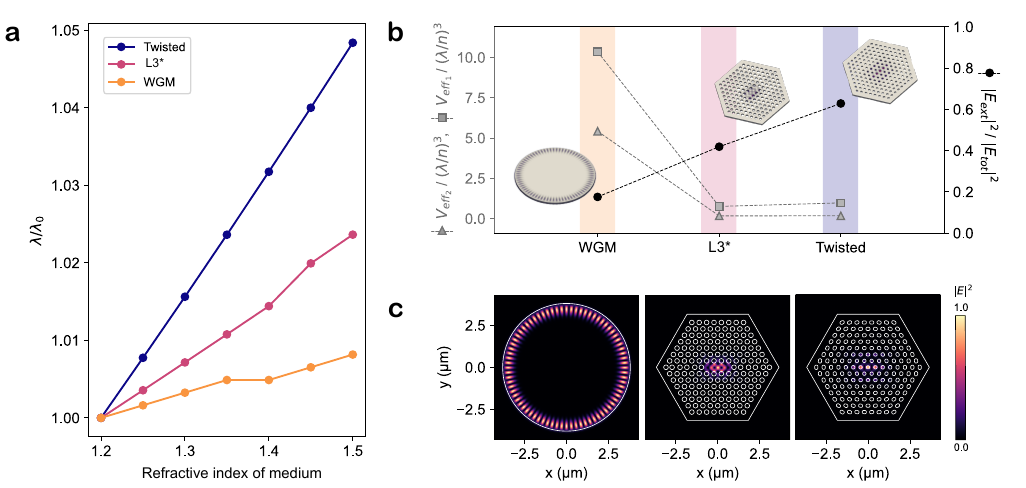}
\caption{\textbf{Improving mode localization and refractive index sensitivity in nanolaser particles by cavity design.} \textbf{(a)} FDTD simulation results showing the change in the mode’s resonant wavelength upon submerging the whispering gallery mode (WGM, yellow), optimized L3 (L3*, purple), and twisted cavity (blue) particles in a surrounding environment of refractive index n = 1.2 to 1.5, a range relevant for sensing in biological environments. \textbf{(b)} Normalized effective mode volumes (left axis, $V_{\text{eff}_1}$ and $V_{\text{eff}_2}$, see Eqs.~\eqref{eq:Veff1} and~\eqref{eq:Veff2}) of the three nanolaser cavities embedded in a medium of n = 1.3, and the ratio of the square of the electric field outside the nanolaser material (e.g. the evanescent field) to the square of the total electric field (circles, right axis). \textbf{(c)} Simulated $|E|^2$ distributions of the WGM, L3* and twisted cavities.}
\label{fig:modevolumes}
\end{figure*}

\section{Biointegration of fully detached PhC nanolasers}
Based on the results of the miniaturization experiments, we determined that a photonic crystal cavity truncated down to only five or six holes on either side of the defect presents a viable design for a standalone nanolaser particle without drastic sacrifices to the lasing performance. As such, hexagonal particles of \SI{7}{\micro\meter} in diameter were produced where we compared our optimized design (Fig.~\ref{fig:devicesize}) with a particle with no hole optimization.

To fabricate nanolasers appropriate for internalization, each hexagonal device was designed with a complete hexagonal trench surrounding it which adds structural support during the detachment from the host wafer (see Fig.~\ref{fig:overview}c and Fig.~\ref{fig:internalized}b).

After hydrochloric acid etching of the InP substrate, particles rested on the InGaAsP etch stop layer (Fig.~\ref{fig:internalized}a). After immediate transferal to an Eppendorf tube of phosphate-buffered saline solution (PBS), brief ultrasonication resulted in complete detachment of particles into the solution (Fig.~\ref{fig:internalized}b), after which the PBS was replaced with cell media in preparation for addition to cell culture.

The nanolaser particles were incubated with NIH3T3 cells for 24 hours to allow sufficient time for internalization. Samples were partitioned for optical characterisation, time-lapse observation, and confocal fluorescence microscopy. To verify nanolaser internalization, a sample of cells was observed continuously for 12 hours under bright-field illumination. Migration of nanolasers within cells was clearly observed (see Fig.~\ref{fig:internalized}d), and the specimen remained viable for several days.

Intracellular nanolasers were interrogated on the same optical setup as used for the measurements of on-chip devices. For optimized nanolasers of \SI{7}{\micro\meter} in diameter, lasing performance was on-par with on-chip devices, with thresholds of 15-30 \unit{\pico\joule} and quality factors up to 13,000 (see Fig.~\ref{fig:internalized}c), approaching the resolution limit of the spectrometer setup. Sustained pumping at above-threshold showed no signs of cell toxicity, consistent with long-term measurements in other intracellular laser platforms~\cite{Martino2019,Schubert2020}.

To evaluate the improvement in performance due to the optimized hole positions, several cells, incubated with regular and optimized nanolasers were interrogated. Although the optimized quality factor fell short of the simulated value, the average quality factor achieved in-cell was about $1.8\times$ that of the regular design (see Ext. Data Fig.~\ref{fig:optsvsreg}).

\section{Demonstrating the versatility of the PhC platform}
The experimentally demonstrated biointegrated L3 nanolasers represent only one of many possibilities enabled by the photonic crystal platform. Taking into account the intended application of intracellular and highly localized sensing, mode designs with significantly increased accessible mode volume and further reduced volume are highly desirable. Having proven the effectiveness of the miniaturized L3 cavity, we trialled one other device design, the \textit{twisted cavity}. Having received recent interest, and having proven a viable option for device truncation~\cite{Ma2023,Wang2025a,Ouyang2024}, the twisted cavity is a superposition of two regular lattices, with a slight relative tilt angle, producing a field maximum in the central air hole, potentially offering significantly increased sensing performance.

We therefore simulated and fabricated a twisted cavity device with dimensions similar to the optimized L3 nanolasers. On-chip lasing showed a quality factor over 6,000 and a further improved lasing threshold compared to the L3 devices (see Fig.~\ref{fig:twisted}). This on-chip demonstration of an alternative cavity design demonstrates the overall versatility of the PhC platform. As new discoveries emerge in the active field of photonics, such as new inverse designs, they could feasibly be adapted and truncated to become intracellular PhC particles, incorporating also other functionalities than the described lasing. 

To demonstrate the difference between existing intracellular nanolasers and the new designs presented by us, we theoretically analysed the sensing performance of the optimized L3 and twisted cavity and compared it with that of a WGM nanodisk laser of the same size. The WGM, L3* (optimized), and twisted designs were simulated in an aqueous environment, showing large differences in the volume that the respective modes occupy (see Fig.~\ref{fig:modevolumes} and Ext. Data Fig.~\ref{fig:modeprofiles}). To quantify the smallness of mode we may refer to its effective mode volume, although several definitions of this quantity exist \cite{Notomi2010}. The most common definition is given by

\begin{equation}
V_{\mathrm{eff}_1}
=
\frac{
 \displaystyle \int \varepsilon(\boldsymbol{r})
 \lvert \boldsymbol{E}(\boldsymbol{r}) \rvert^2 \, d^3\boldsymbol{r}
}{
 \displaystyle \max \left[
  \varepsilon(\boldsymbol{r})
  \lvert \boldsymbol{E}(\boldsymbol{r}) \rvert^2
 \right] .
}
\label{eq:Veff1}
\end{equation}

Another useful definition, representing the three-dimensional full-width-half-maximum of the mode, is given by

\begin{equation}
V_{\mathrm{eff}_2}
=
\int_{\Omega} d^3 \boldsymbol{r},
\,\,
\Omega
=
\left\{
 \boldsymbol{r}
 \;:\;
 \lvert \boldsymbol{E}(\boldsymbol{r}) \rvert^2
 >
 \frac{1}{2}
 \max \left[
  \lvert \boldsymbol{E}(\boldsymbol{r}) \rvert^2
 \right]
\right\} .
\label{eq:Veff2}
\end{equation}

In Fig.~\ref{fig:modevolumes}b we see the calculated mode volumes for the three cavity types. Evidently, the photonic crystal devices express mode volumes that are almost an order of magnitude smaller compared to that of a WGM nanodisk. We also find that our optimized L3 cavity has the smallest mode volume of only 0.76 ($\lambda$ / $n$)$^3$, compared to a volume of 10.35 ($\lambda$ / $n$)$^3$ for the WGM mode, displaying the extreme light confinement that can be achieved in a PhC cavity.

However, it is also interesting to compare the fraction of mode volume which is evanescent to the device, that is, which does not reside within the semiconductor, but in the surrounding medium (e.g. cytosol). Most sensing regimes rely upon the target of interest coupling to the evanescent field of the resonator, which usually extends no more than a few tens of nanometers from the surface. For example, the environment immediately surrounding a nanolaser may temporarily or permanently rise in refractive index due to protein binding, the transient proximity of an organelle, or a change in the concentration of a certain biomarker. To simulate this effect, we varied the refractive index (RI) of the surrounding medium in the simulation and found that the resonant wavelength response from the cavities correlates strongly with the degree of field externalization (see Fig.~\ref{fig:modevolumes}a). Here, the twisted cavity shows a refractive index sensitivity six times that of the WGM cavity, despite having an order of magnitude smaller mode volume. This result holds promise for improving upon the sensitivities achieved by intracellular WGM sensing experiments \cite{Thomson2025, Schubert2020}.

\section{Conclusions and Outlook}

Here, we propose and demonstrate the miniaturization of functional photonic crystal arrays down to dimensions that make them suitable for integration into single biological cells. Our theoretical and experimental optimization identified a minimum array size to generate PhC nanolasers, reaching a surface area as small as \SI{30}{\micro\meter\squared}. We believe that miniaturisation of PhC will allow us to answer more complex biological questions using structurally and functionally targeted intracellular probes. As an example, we created lasing probes with sensing volumes on the order of tens of attolitres and successfully internalized them into living cells. The lasing signal received from within live cells was barely degraded by the cellular environment and cells remained visually unaffected by the lasing action.

Thanks to an abundance of active research into manipulating the spectral, spatial, and temporal features of photonic crystals, and the availability of a myriad of structural parameters, the minimal-footprint PhC nanolaser particle has the potential for various other designs than those presented in this paper. Our nanofabrication procedure can be clearly extended to include additional plasmonic, chemical, or biological elements, similar to those employed in existing extracellular sensing devices.

By enabling real-time, non-destructive interrogation of the intracellular environment with exceptional spatial selectivity and sensitivity, this platform opens new opportunities to investigate fundamental biological processes, and clinically relevant phenomena such as protein activity and dynamic drug response at the single-cell level.

\begin{backmatter}

\bmsection{Funding} This work received financial support from the European Research Council under the European Union’s Horizon Europe Framework Program/ERC Starting Grant agreement no. 101043047 (HYPERION, to M.S.). Instrument funding by the Deutsche Forschungsgemeinschaft in cooperation with the Ministerium für Kunst und Wissenschaft of North Rhine-Westphalia (INST 216/1121-1 FUGG, 470088514; INST 216/1120-1 FUGG, 469988234).

\bmsection{Author contributions} CAT designed, fabricated, characterized and optimized all nanolaser devices, and performed optical simulations and calculations. AS designed and built the optical setup. NP and CAT performed cell experiments and NP performed confocal microscopy imaging. VD assisted in device design and fabrication optimization. MS conceived and supervised the project and acquired funding. CAT and MS wrote the manuscript with input from all authors. 

\bmsection{Acknowledgment} The authors are grateful to Wendong Tan and Valerie Tan for their helpful contributions towards the development of the experimental processes.
\bmsection{Disclosures} The authors declare no conflicts of interest.
\end{backmatter}
\bibliography{InGaAsPPhCBibPreprint}

\clearpage

\renewcommand{\figurename}{Extended Data Fig}
\setcounter{figure}{0}
\renewcommand{\thefigure}{\arabic{figure}}
\captionsetup[figure]{labelsep=period}

\begin{figure*}[p]
\centering
\includegraphics[width=\linewidth]{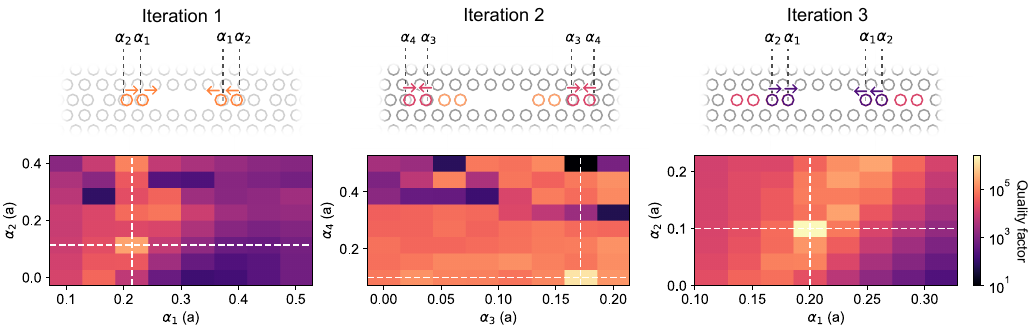}
\caption{\textbf{Iterative grid search optimization for the positions of holes adjacent to the defect cavity, to produce a cavity of high quality factor (Q).} Instead of starting the optimization from a regular L3 lattice, a five-hole optimization for an L3 cavity performed by Minkov et. al for a silicon resonator around 1.55 \unit{\micro\meter} was used as the basis~\cite{Minkov2014}. Here, the five adjacent holes were offset from the regular lattice by $\alpha_1 = 0.337a$, $\alpha_2 = 0.270a$, $\alpha_3 = 0.088a$, $\alpha_4 = 0.323a$, and $\alpha_5 = 0.173a$, where $a$ is the lattice spacing. Our simulated device had a lattice spacing of 440 nm, a hole radius of $0.32a$ and a slab thickness of 240 nm, producing a fundamental L3 mode in InGaAsP close to the target wavelength of 1590 nm. In the first iteration, we performed a sweep of the offsets for the first and second holes adjacent to the defect, centred around the basis design. The second iteration used the optimal $\alpha_1$ and $\alpha_2$ found from the first iteration, and swept over offsets for the third and fourth holes. The final iteration again swept over $\alpha_1$ and $\alpha_2$ in a smaller range, producing a final theoretical Q maximum of 2,222,171. The value of $\alpha_5$ was not swept in order to minimize simulation time.}
\label{fig:optimization}
\end{figure*}

\begin{figure*}[p]
\centering
\includegraphics[width=\linewidth]{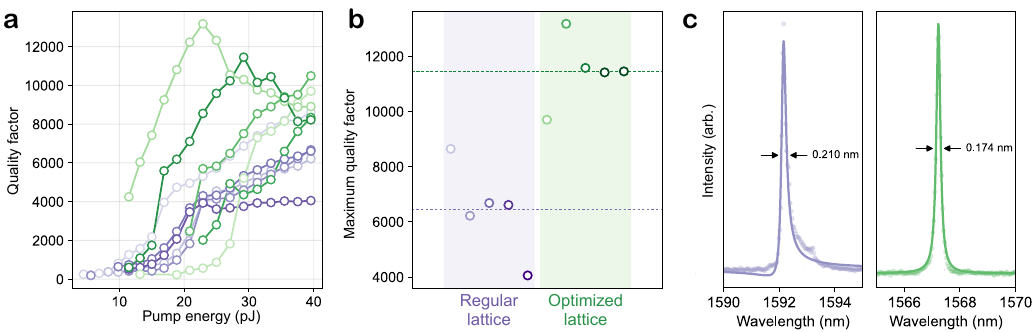}
\caption{\textbf{Laser performance of intracellular lasers with optimized L3 cavities or L3 cavities with a regular hexagonal lattice.} \textbf{(a)} Quality factors obtained from measured laser thresholds of nanolasers located inside live fibroblast cells with either the optimized lattice (green) or the regular lattice (purple). The regular design had a lattice spacing of 440 nm and a hexagon side length of 3.5 \unit{\micro\meter}. The optimized design had a lattice spacing of 420 nm and a side length of 3.6 \unit{\micro\meter}. Slight adjustments in the size parameters were necessary to match the resonant wavelengths of the two designs. \textbf{(b)} An overview of the maximum quality factors of the measured nanolasers. \textbf{(c)} A comparison of the laser spectrum between the highest quality factor regular lattice nanolaser (purple) and the highest quality factor optimized lattice nanolaser (green). }
\label{fig:optsvsreg}
\end{figure*}

\begin{figure*}[p] 
\centering
\includegraphics[width=0.5\linewidth]{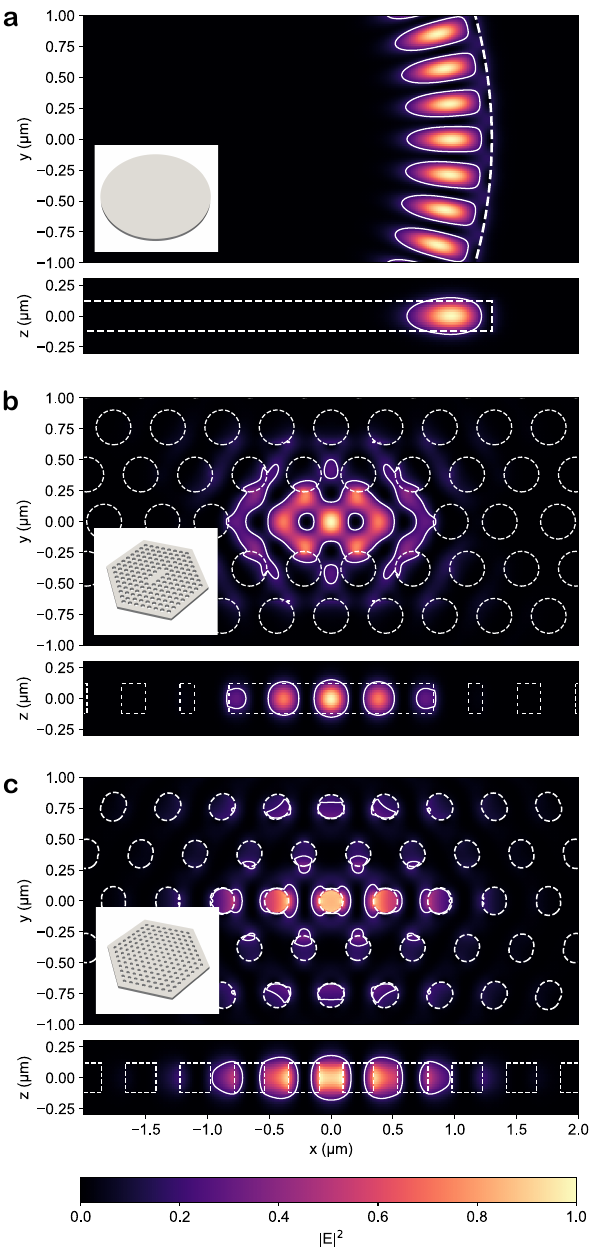} 
\caption{\textbf{FDTD simulated mode profiles for the WGM (a), optimized L3 (b), and twisted (c) cavity designs.} Iso-contours (solid lines) at 20~\% of the $|E|^2$ field maximum (top), displaying differing levels of mode localization and confinement to the semiconductor material. x-z cross-sections (bottom) show the extent of the evanescent field differs between the cavity types. Dashed lines represent the boundary between the semiconductor material and the surrounding medium.}
\label{fig:modeprofiles}
\end{figure*}

\begin{figure*}[p]
\centering
\includegraphics[width=0.5\linewidth]{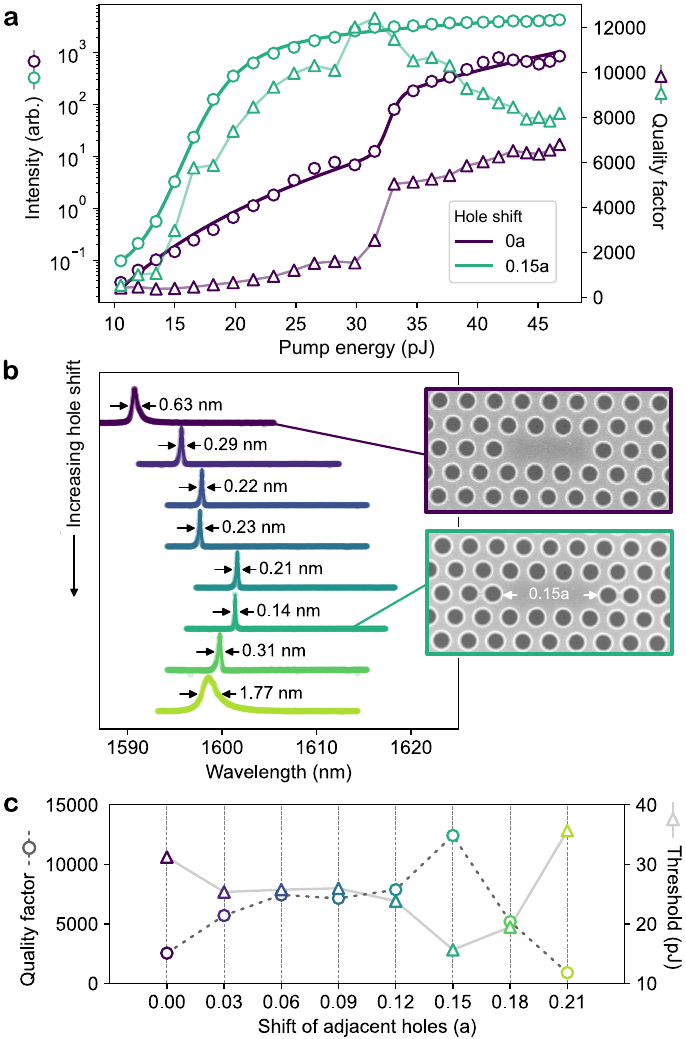}
\caption{\textbf{Investigation into minimally changing the regular lattice of a L3 cavity by shifting only the first holes adjacent to the defect.} \textbf{(a)} Emitted power (left axis) and quality factor (right axis) of a regular (purple) and optimized (green) nanolaser. Devices (on-chip) with 10 hole pairs adjacent to the defect (i.e. a \SI{10}{\micro\meter} array diameter) and with a regular lattice of lattice spacing 440 nm, and hole radius $0.32a$ are investigated. \textbf{(b)} Lasing spectra for a series of nanolasers obtained by linearly shifting the first adjacent hole pair from 0 to $0.21a$. The maximum quality factor and lowest threshold is found at a hole shift of $0.15a$, consistent with previous experiments~\cite{Akahane2003}. All spectra were measured at a pump pulse energy of \SI{31.5}{\pico\joule}. \textbf{(c)} An overview of the thresholds, and quality factors achieved at a pump energy of \SI{31.5}{\pico\joule} (as in (b)) for the eight investigated cavities.}
\label{fig:twoshifted}
\end{figure*}

\renewcommand{\figurename}{Fig.} 
\renewcommand{\thefigure}{\arabic{figure}}

\end{document}